\documentclass{nature}

\usepackage{epstopdf}
\usepackage{url}

\usepackage{graphicx}
\usepackage{subfigure} 
\usepackage{placeins}
 
\usepackage{flushend}

\usepackage{algorithm}
\usepackage{algorithmic}
\usepackage{amsmath}
\usepackage{amssymb}
\usepackage{todonotes}
\usepackage{booktabs}

\usepackage{color}
\newcommand{\mum}{\mu {\rm m}}
\definecolor{willcolor}{RGB}{23, 127, 117}
\definecolor{evacolor}{RGB}{23, 127, 117}
\definecolor{bcolor}{RGB}{182, 33, 45}

\newcommand{\website}{\url{docs.neurodata.io/xbrain}}
\title{Quantifying mesoscale neuroanatomy using X-ray microtomography}

\author{Eva L. Dyer$^{1,2}$, William Gray Roncal$^{3,4}$, Hugo L. Fernandes$^{1,2}$, Doga G\"ursoy$^{5}$, Vincent De Andrade$^{5}$, Rafael Vescovi$^{5}$, Kamel Fezzaa$^{5}$, Xianghui Xiao$^{5}$, Joshua T. Vogelstein$^{6,7}$,
Chris Jacobsen$^{5,8}$, \\Konrad P. K\"ording$^{1,2,*}$ \& Narayanan Kasthuri$^{9,10,*}$}

\begin{document}
\maketitle

\begin{affiliations}
 \item Dept. of Physical Medicine and Rehabilitation, Northwestern University, Chicago, IL
\item Sensory Motor Performance Program, Rehabilitation Institute of Chicago, Chicago, IL
\item The Johns Hopkins University Applied Physics Laboratory, Laurel, MD
\item Dept. of Computer Science, The Johns Hopkins University, Baltimore, MD
\item Advanced Photon Source, Argonne National Laboratory, Lemont, IL
\item Dept. of Biomedical Engineering, The Johns Hopkins University, Baltimore, MD
\item Institute of Computational Medicine, The Johns Hopkins University, Baltimore, MD
\item Dept. of Physics and Astronomy, Northwestern University, Evanston, IL
\item Center for Nanoscale Materials, Argonne National Laboratory, Lemont, IL
\item Dept. of Neurobiology, University of Chicago, Chicago, IL \\
 * Contributed equally to this paper
\end{affiliations}
 
\begin{abstract}
Methods for resolving the 3D microstructure of the brain typically start by thinly slicing and staining the brain, and then imaging each individual section with visible light photons or electrons. In contrast, X-rays can be used to image thick samples, providing a rapid approach for producing large 3D brain maps without sectioning. Here we demonstrate the use of synchrotron X-ray microtomography ($\mu$CT) for producing mesoscale ($1~\mum^3$ resolution) brain maps from millimeter-scale volumes of mouse brain. We introduce a pipeline for $\mu$CT-based brain mapping that combines methods for sample preparation, imaging, automated segmentation of image volumes into cells and blood vessels, and statistical analysis of the resulting brain structures. Our results demonstrate that X-ray tomography promises rapid quantification of large brain volumes, complementing other brain mapping and connectomics efforts.
\end{abstract}
 
\vspace{2mm}
Large-scale brain maps that provide a glimpse into the cellular and vascular architecture of the brain are essential for understanding neuroanatomy, and its relation to function and disease\cite{lichtman2011big}. Unfortunately, acquiring high resolution brain maps is still difficult and time intensive\cite{economo2016platform}. Conventional light and electron microscopy (EM) methods require sectioning tissue into thin slices ($\mu$m scale), imaging each slice individually, and then stitching the images back together to get a 3D brain map. For example, stitching BigBrain---a 3D reconstruction of a full human brain at $20$ $\mu{\rm m}$ isotropic resolution---required approximately $1,000$ hours to complete\cite{amunts2013bigbrain}. Electron scatter occurs at even smaller depths than visible light and as a consequence, EM requires even thinner slices ($\sim 30$~nm). It thus takes approximately three months to image a cubic millimeter of brain tissue at $20$ nm resolution\cite{eberle2015high}, requiring approximately two petabytes on disk. Thus, methods for quickly imaging the brain's microstructure are critical for understanding and comparing the structure and function of many brains.

Tissue clearing approaches such as CLARITY\cite{chung2013clarity} and expansion microscopy\cite{chen2015expansion} address some of the challenges associated with large samples. However, unlike EM, these techniques produce sparse reconstructions which reveal only subsets of neurons in the volume. In addition, tissue clearing requires the removal of scattering membranes in tissue samples, which renders them incompatible with subsequent serial section electron microscopy to identify individual neuronal connections. As a consequence, interrogation of the sample is primarily limited to the mesoscale and it is challenging to re-investigate the same tissue at higher resolution. Therefore, new approaches, capable of producing large-scale complete mesoscale reconstructions of the brain, are required.


X-ray microtomography ($\mu$CT) provides a unique and largely untapped opportunity for brain mapping. X-rays can theoretically penetrate through centimeter-scale brain volumes with micron resolution, without the need for sectioning. Recent studies have demonstrated the utility of benchtop $\mu$CT systems for neuroscience\cite{bushong2015x,mikula2015high}. However, using benchtop systems for large-scale brain mapping efforts is difficult due to the long exposure times needed to collect even a single image---imaging a cubic mm brain sample at $1 \mum$ resolution would take at least 13 hours on state-of-the-art scanners\cite{arillo_sysent_2015}. Fortunately, synchrotron-based $\mu$CT offers far higher photon flux and thus provides an avenue for the rapid acquisition (two orders of magnitude speedup) of large brain volumes\cite{mizutani2009microtomographic, mizutani2010unveiling, mizutani2012x}. However, $\mu$CT has not yet been adapted to meet the demands of large-scale brain mapping efforts. 

Here we introduce a pipeline for quantifying mesoscale neuroanatomy with $\mu$CT. We demonstrate that samples  fixed with aldehydes, stained with osmium, and embedded in plastic can be imaged with high-energy synchrotron radiation. The resulting image datasets provide sufficient isotropic resolution ($1~\mum^3$) and contrast to resolve the 3D structure of neuronal and glial cell bodies, vasculature, and segments of large apical dendrites and myelinated axons. We can subsequently section the same samples and image them with an electron microscope; the result shows excellent preservation of the ultrastructure and straightforward correspondence (leading to easy co-registration) between X-ray and EM datasets. These results confirm that $\mu$CT can be used to produce imaging data with sufficient resolution to compute mesoscale brain maps containing information about the cyto- and myelo-architecture of cortex. We developed a suite of open-source tools, called X-BRAIN (X-ray Brain Reconstruction, Analytics and Inference for Neuroanatomy) (\website) for cell detection, blood vessel segmentation, and statistical analyses of X-ray image volumes. $\mu$CT in combination with image parsing techniques offers an effective path from brain specimens to mesoscale brain maps. 

\section{Results}
\begin{figure}[t!]
\centering{
\includegraphics[width=\textwidth]{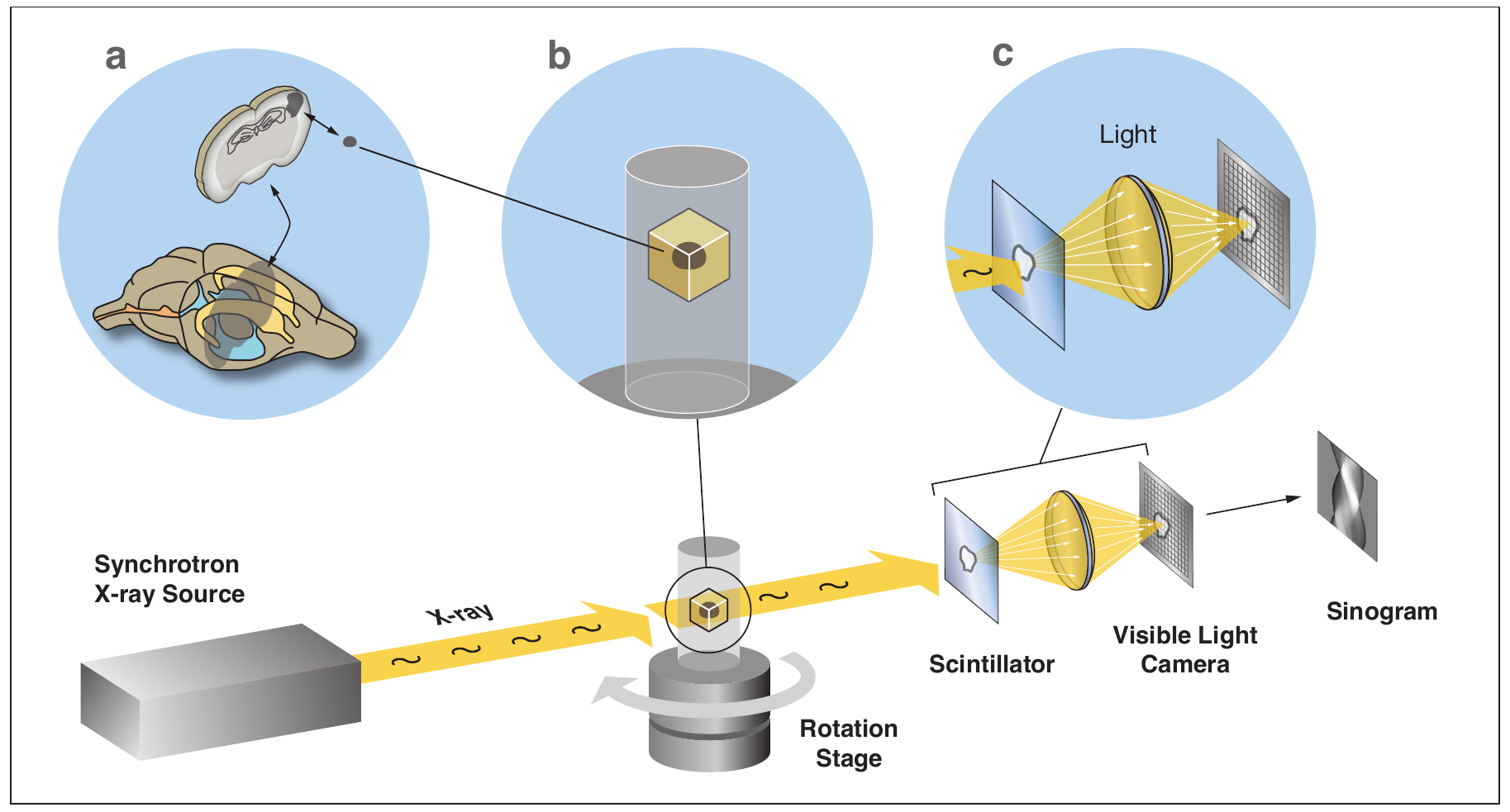}
\caption{{\em Synchrotron X-ray imaging of millimeter-sized brain volumes.} A schematic of our sample preparation and imaging setup are displayed along the bottom: from left to right, we show the synchrotron X-ray source interacting with a embedded sample of brain tissue as it is rotated to collect multi-angle projections. To collect projection data, X-rays are passed through a scintillator crystal which converts X-rays into visible light photons, and then focused onto a light camera sensor. Finally, we obtain a sinogram from the sample by collecting data from a row of sensor pixels. Above, we show a more detailed depiction of the (a) sample preparation, (b) sample mounted in the instrument, and (c) conversion and focusing of X-rays to light photons. \label{fig:fig1}  }
}
\end{figure}

\subsection{X-ray tomography on a millimeter- scale brain sample.} 
Using the 2-BM synchrotron beamline at the Advanced Photon Source (APS)\cite{de2014scientific}, we obtained tomography data from cubic mm volumes of brain tissue  (Fig.\ \ref{fig:fig1}). Samples were fixed with aldehydes, stained with osmium, and embedded in plastic, making them compatible with subsequent large volume EM\cite{tapia2012high}.
Stacks of projection images were acquired by rotating the sample at ($3$,$000$) uniformly spaced rotation angles from 0 to 180 degrees and measuring the propagation of X-rays through the sample at each rotation angle (Fig.~\ref{fig:fig1}b). Radiographs are recorded with an indirect detection system consisting of a thin scintillator which converts the transmitted X-rays into visible light (Fig.\ \ref{fig:fig1}c). The light is then focused by an objective lens on a charged coupled detector (CCD) array, producing images with equivalent pixel size of ($0.65 \mum$)$^2$ at the sample plane. After calibrating the instrument, collecting the main dataset studied in this paper (10.6 Gigavoxels) took approximately six minutes. To obtain high contrast images, data acquisition was performed in propagation-based phase contrast mode by increasing the distance between the detector and sample to several tens of centimeters with a pink beam ($\Delta$E/E = $10^{-2}$) set respectively to 30 keV. To reconstruct a 3D image volume from the projection data, phase retrieval was performed on each projection using the well established Paganin algorithm\cite{paganin2002simultaneous}, followed by volume reconstruction using the open source TomoPy package\cite{gursoy2014tomopy}. The resulting image volume provides the data for our segmentation and analysis methods.

\begin{figure}
\centering{
\includegraphics[width=\textwidth]{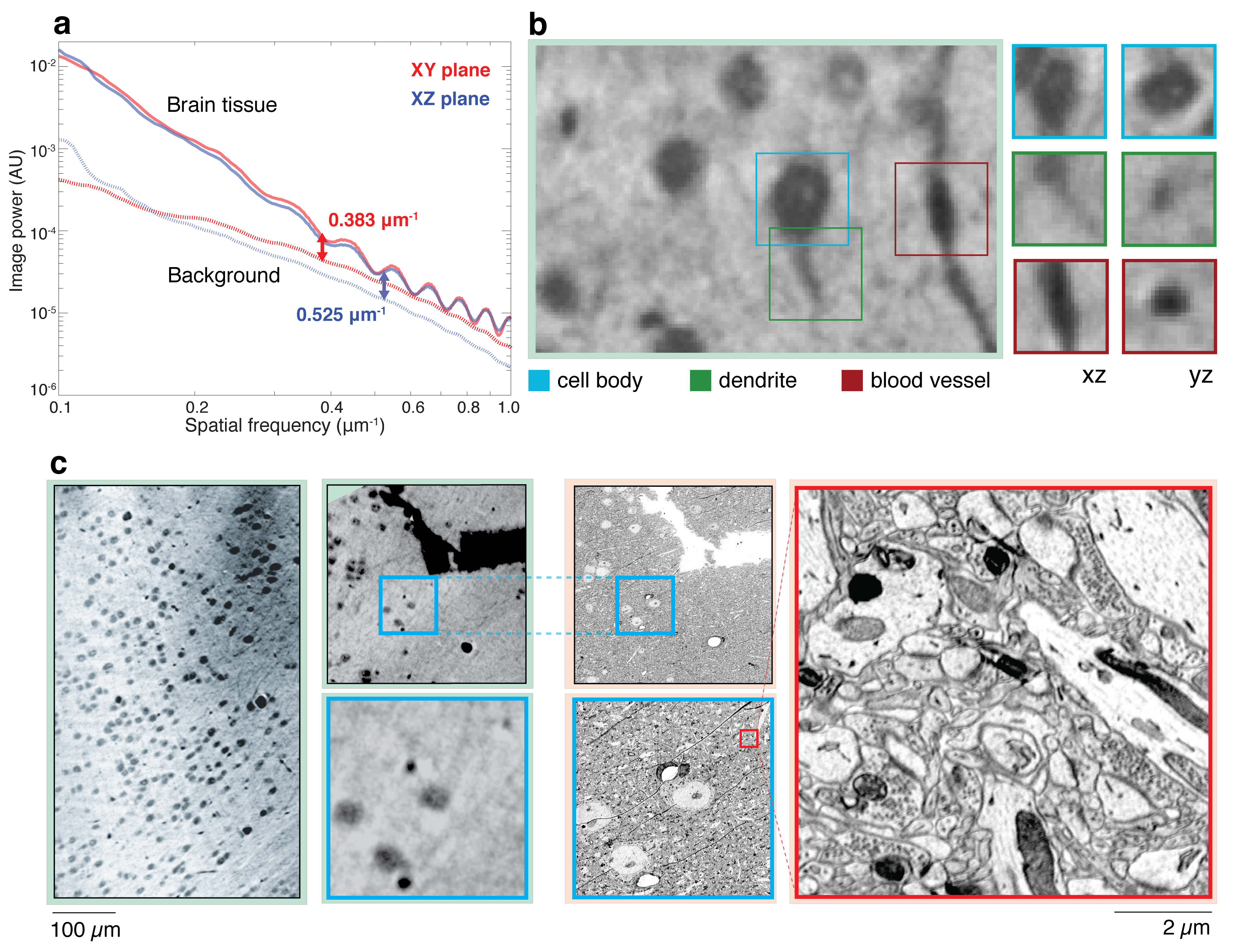}
\caption{\small \linespread{0.2}{\em Synchrotron X-ray imaging provides micron resolution of brain volumes.} (a) From our reconstructed volumes, we compared two regions: one containing brain tissue (which we denote as signal power spectra, or SPS), and a background region without brain tissue (which we denote as noise power spectra, or NPS). We display the signal and noise power spectra (SPS and NPS) computed by taking a series of 256 XY or transverse planes and vertical planes (virtual slices) and averaging their power spectra to measure the resolution parallel to and perpendicular to the rotation axis. We then fit a second order polynomial to the SPS to account for artifacts introduced during phase retrieval. When measuring the gap between the smoothed power spectra and the background spectra, the SPS is 5 times the noise (following the Rose criterion\cite{rose_jsmpe_1946} for detectability) at a spatial frequency of $0.383~\mu$m$^{-1}$ in XY, and $0.525~\mu$m$^{-1}$ in XZ. This indicates a half-period spatial resolution of 1.31 $\mu$m in XY and 0.95 $\mu$m in XZ. (b) We show multi-view projections of X-ray image volumes, where the 3D structure of cells, vessels, and dendrites is visible. (c) We show $\mu$CT and EM images of the same sample, collected at three different pixel sizes ($0.65~\mum$, $100~{\rm nm}$, $3~{\rm nm}$). On the left, we display a subset of a single virtual slice from an X-ray tomogram which spans multiple layers of mouse somatosensory cortex ($0.65~\mum$ pixel size). Next to this, we show a subset of the same image in (a) which highlights a configuration of three cell bodies with distinct micro-vessels nearby (outlined in blue). This tissue block was subsequently serial-sectioned and imaged in a scanning electron microscope with $100$ nm and $3$ nm pixel sizes. We located the same configuration of cells in the EM dataset (outlined in blue) and observe that the EM ultrastructure is well preserved after $\mu$CT (right in red).\label{fig:fig2}}
}
\end{figure}

To quantify the resolution of our reconstructed X-ray image volumes, we obtained digitally vignetted sub-fields from regions with brain tissue (signal), and without (background) and computed their respective 2D Fourier power spectra.  The signal power spectra (SPS) for regions with brain tissue is five times larger than in the noise power spectra (NPS) computed in background regions at a half-period resolution of about 1.31 $\mu$m in the XY or transverse plane, and 0.95 $\mu$m in the XZ or vertical plane (Fig.~ \ref{fig:fig2}a).  
This near-isotropic spatial resolution simplifies data analysis when compared to other imaging approaches that may give very non-isotropic values of spatial resolution in XY versus XZ.

X-ray images allow resolving the putative location and morphology of cell bodies, blood vessels, and segments of large neurites (Fig.\ \ref{fig:fig2}b, Supp.\ Fig.\ \ref{suppfig:axons}). We estimate that voxels inside cells are on average $4.56\pm 1.13$ dB (mean$\pm$std) brighter then their immediate surround (see Methods). At this contrast level, it is possible to discern the location and size of cells in the sample. Blood vessels are also visible in the sample and provide even stronger contrast than cell bodies, making them much easier to track. This signal strength suggests that we should be able to segment the tissue into cell bodies and blood vessels, which we validate with our automated techniques.

After collecting $\mu$CT data, we performed ultra-thin sectioning and electron microscopic imaging of the same sample. EM confirmed the identity of the cell bodies, myelinated axons, and blood vessels, corresponding to those annotated in the $\mu$CT dataset (Fig.~\ref{fig:fig2}c), suggesting that  details seen in the X-ray dataset are bona fide and not spurious results of our imaging and processing pipelines. In addition, we noticed no changes in the microtome sectioning properties of the epon-embedded brain tissue, nor did we see any obvious signs of irradiation-induced structural damage in the scanning electron micrographs obtained from these sections.  Structures like synapses and mitochondria are still clearly evident (Fig.~\ref{fig:fig2}c). This is consistent with our calculated radiation dose of about 3 kGy during the collection of the X-ray tomography data; this dose is well below the dose affecting the dissolution rate of radiation-sensitive polymers like poly(methyl methacrylate) or PMMA\cite{zhang_jvstb_1995} (1000 kGy), and the  dose at which glutaraldehyde-fixed wet chromosomes start to show mass loss\cite{williams_jm_1993}(70 MGy). Our results confirm that $\mu$CT and EM can  be coupled to produce multi-resolution brain maps.  

A good dataset, at minimum, should allow human annotators to clearly see the structures of interest and in turn, reliably annotate them. We thus measured human annotator ability in finding and labeling cell bodies and blood vessels in multi-view projections (orthogonal 2D projection planes) of the 3D image data. Two expert annotators (A0 and A1) were instructed to label the boundaries of all of the cells and vessels in a small volume $(100~\mum)^3$ of X-ray image data with ITK-Snap\cite{py06nimg}. When provided 3D context, pixel-level agreement between annotators of the cell bodies and blood vessels were (precision, recall) $(p,r) = (0.835,0.76)$ and $(p,r) = (0.85,0.73)$, for V2 and V3 respectively (see Methods). We further measured the discrepancy between the centroids of cell bodies across both annotators and find nearly perfect agreement $(p,r) = (1,0.989)$. 
While precise manual segmentation of the boundaries of cell bodies and vessels is challenging, we observe that the object-level agreement between annotations of the centers of cell bodies is high, showing that performance on the detection task is nearly identical. We conclude that the data is of sufficient quality to segment cell bodies and vessels with automated methods.


\begin{figure}
\centering{
\includegraphics[width=0.8\textwidth]{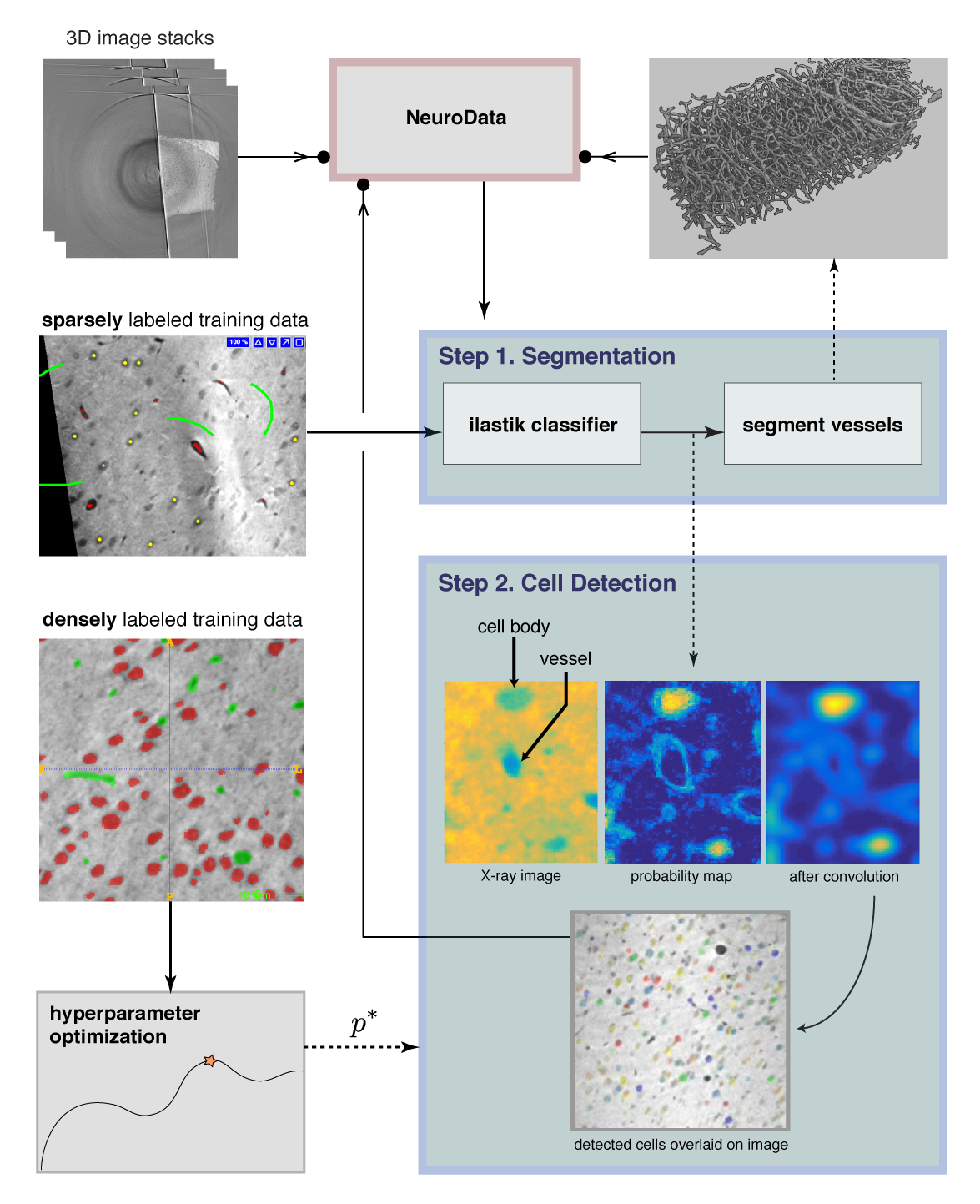}
\vspace{-2mm}
\caption{\small {\em Image processing and computer vision pipeline for segmentation and cell detection.} A block diagram of the entire workflow is displayed. We show the integration of sparsely labeled training data into our segmentation module (Step 1) to train a random forest classifier using ilastik. We use densely annotated training data to perform hyperparameter optimization to tune our cell detection algorithm (Step 2). The final map of detected cells is displayed at bottom of Step 2, with detected cells overlaid on top of the original X-ray image. Solid arrows indicate inputs into a module, outputs are indicated by dashed arrows, and outputs that are stored in NeuroData are indicated with a filled circle terminal. \label{fig:fig3}}
}
\end{figure}

\subsection{Automated methods for segmentation and cell detection in X-ray volumes.}
The datasets afforded by X-ray tomography are too large to be analyzed by humans. Therefore, we developed automatic 3D segmentation algorithms to extract cells and vessels from the image volumes. We created a suite of tools for extracting and visualizing mesoscale maps from 3D stacks of X-ray images (Fig.\ \ref{fig:fig3}). This set of tools, X-BRAIN ({\em X-ray Brain Reconstruction, Analytics, and Inference for Neuroanatomy}), consists of image processing and computer vision methods for preprocessing and artifact removal, segmentation,  estimating the location and size of cells, and vessel segmentation. We also provide methods for large-scale analyses of these data to compute relevant statistics on the reconstructed maps of the cells and vessels. X-BRAIN is implemented in Matlab and Python and both code and data are openly available through \website, providing a community resource for the automated segmentation and quantification of mesoscale brain anatomy.

Our main image processing and computer vision pipeline (Step 1-2 in Fig.\ \ref{fig:fig3}) consists of methods for segmenting blood vessels and detecting the location and size of cells in the volume. In the initial step of our workflow, we train a classifier to predict the probability that each brain voxel belongs to each of the three classes: cell body, blood vessel, and background (other). To do this, we use a tool called ilastik to sparsely annotate data and build a random forest classifier using intensity, edge, and gradient features computed on the image volume\cite{ilastik}. This classification procedure returns three probability maps $\mathcal{P} = \{ P_c, P_v, P_{bg}\}$, which collectively provide the probability tuple $p(x,y,z) = \{ P_c(x,y,z), P_v(x,y,z), P_{bg}(x,y,z)\}$ that each voxel, whose position is denoted by $(x,y,z)$, is a cell, vessel, or lies in the background (output of ilastik in Step 1 of Fig.\ \ref{fig:fig3}). This classification procedure provides an easy and intuitive way to provide an estimate of which voxels correspond to cell bodies and blood vessels. 

The simplest way to convert a probability map to a (binary) segmentation, is to threshold the probabilities and group the resulting structures that pass this test into connected components.  In the case of vessel segmentation, we can employ this procedure with minimal tweaks. To segment vessels in the sample, we threshold the vessel probability map and then apply simple morphological filtering operations to clean and smooth the resulting binary data (see Methods). Visual inspection and subsequent quantification of precision and recall of vessel segmentation suggests a high-degree of accuracy through this simple post-processing of the ilastik outputs. 

Applying the same thresholding procedure used for vessel segmentation, to the segmentation of cells, is difficult because neurons and blood vessels are often densely packed. In this case, simple thresholding-based approaches tend to group clusters of cells and vessels together (see Supp.\ Materials Fig. \ref{fig:algofig}). We thus developed an algorithm for cell detection (Step 2 in Fig.\ \ref{fig:fig3}), which produces estimates of the centroids and radii of detected cells. Our method iteratively selects a new candidate centroid based upon the correlation between the probability map and a (fixed radius) spherical template  which serves to enforce our biologically-inspired shape prior. We use a frequency-based approach to convolve a spherical template with the cell probability map and greedily select ``hotspots'' which are likely to contain cell bodies (see Methods for further details). Our method  leverages prior knowledge of the approximate size and spherical shape of cells to select sphere-like objects from the pre-filtered probabilities to resolve situations where neurons and blood vessels appear in close proximity.

After finding the centroids of all detected cells, we can then efficiently estimate their sizes. To do this, we center a small spherical template at the detected center of each cell and estimate the cell size by varying the template size.  When the template can no longer be inscribed within the cell body, we observe a sharp decay in the correlation. Thus, we compute the correlation between the probability map while increasing the diameter of the spherical template, find the maximum decrease in correlation, and select this diameter as our estimate of the cell size. This operation has low complexity and can be performed on the entire (cubic mm) dataset on a single workstation. Once we have detected cells, estimating the diameter of the cell body is a simple one-dimensional fitting problem.

\begin{figure}
\centering{
\includegraphics[width=\textwidth]{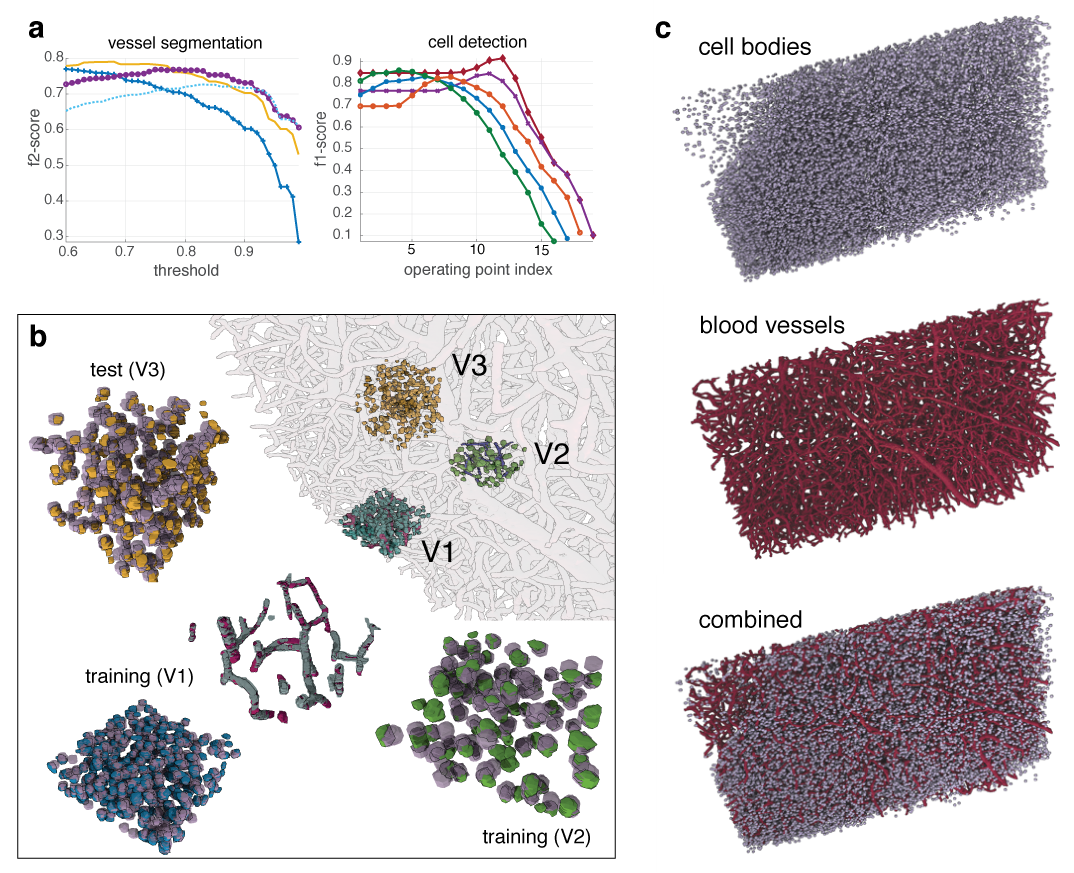}
\vspace{-2mm}
\caption{ \small { \em Automatic methods for segmentation and cell detection reveal dense mesoscale brain maps.} In (a), we display the performance of our vessel segmentation (left) and cell detection (right) methods as we vary different hyperparameters that affect the performance of the method. To optimize performance of our vessel segmentation method, we compute the $f_2$ score --- emphasizing recall --- for multiple operating points (each curve represents a fixed parameter set with a varying vessel segmentation threshold).  To measure performance for cell detection, we compute the $f_1$ score --- balancing precision and recall --- for multiple operating points (curves) as we increase the stopping criterion (x-axis) in our greedy cell finder algorithm. In (b), we show the results of our cell detection and vessel segmentation algorithms on manually annotated test datasets. The training (V1, V2) and test (V3) volumes are visualized, both inside the entire volume (right) and individually (left). We overlay the results of X-BRAIN on the three volumes, based upon the best operating point selected by our parameter optimization approach in (a). In (c), we show  renderings of the output of our cell detection and vessel segmentation algorithms on the entire cubic mm sample. \label{fig:fig4}
}}
\end{figure}

To optimize each stage of our segmentation pipeline, we performed an exhaustive grid search to find the set of hyperparameters (i.e., threshold parameters for cell/vessel detection, the size of spherical template, and the stopping criterion for the cell finder) that maximize a combination of the precision and recall ($f$-score) between our algorithm's output and manually annotated data from volume V1 (Fig.\ \ref{fig:fig4}a-b). After tuning our cell detection algorithm to find the best set of hyperparameters, we obtained a precision and recall of $(p,r) = (0.86,0.84)$ on the same volume. Our initial results on this training volume and visual inspection of large-scale runs (Fig.\ \ref{fig:fig4}c) suggest that our methods provide reliable maps of the cells and vessels in the sample.

The image data varies across space, due to various details of the imaging and reconstruction pipeline. Therefore, it is important to test that our segmentation algorithm works reliably across regions previously unseen during classifier training. We thus labeled and tested our cell detection algorithm on two additional test cubes V2 and V3 (Fig.\ \ref{fig:fig4}b) that are spatially disjoint from V1 and each other. V2 served an initial test set, as we added some sparse training data from this volume to train our ilastik classifier. V3 served as a held-out test set, as the location of this cube was unknown before tuning and running the algorithm on the entire dataset. After obtaining ground truth labels, we ran X-BRAIN on V2 and V3, using the set of parameters selected by optimizing our method on V1. The precision and recall is given by $(p,r) =(0.83,0.76)$ and $(p,r) = (0.94,0.78)$, for V2 and V3 respectively. These results suggest that X-BRAIN generalizes well across different regions of the sample, and is robust to fluctuations in brightness and contrast.

\begin{figure}
\centering{
\includegraphics[width=0.85\textwidth]{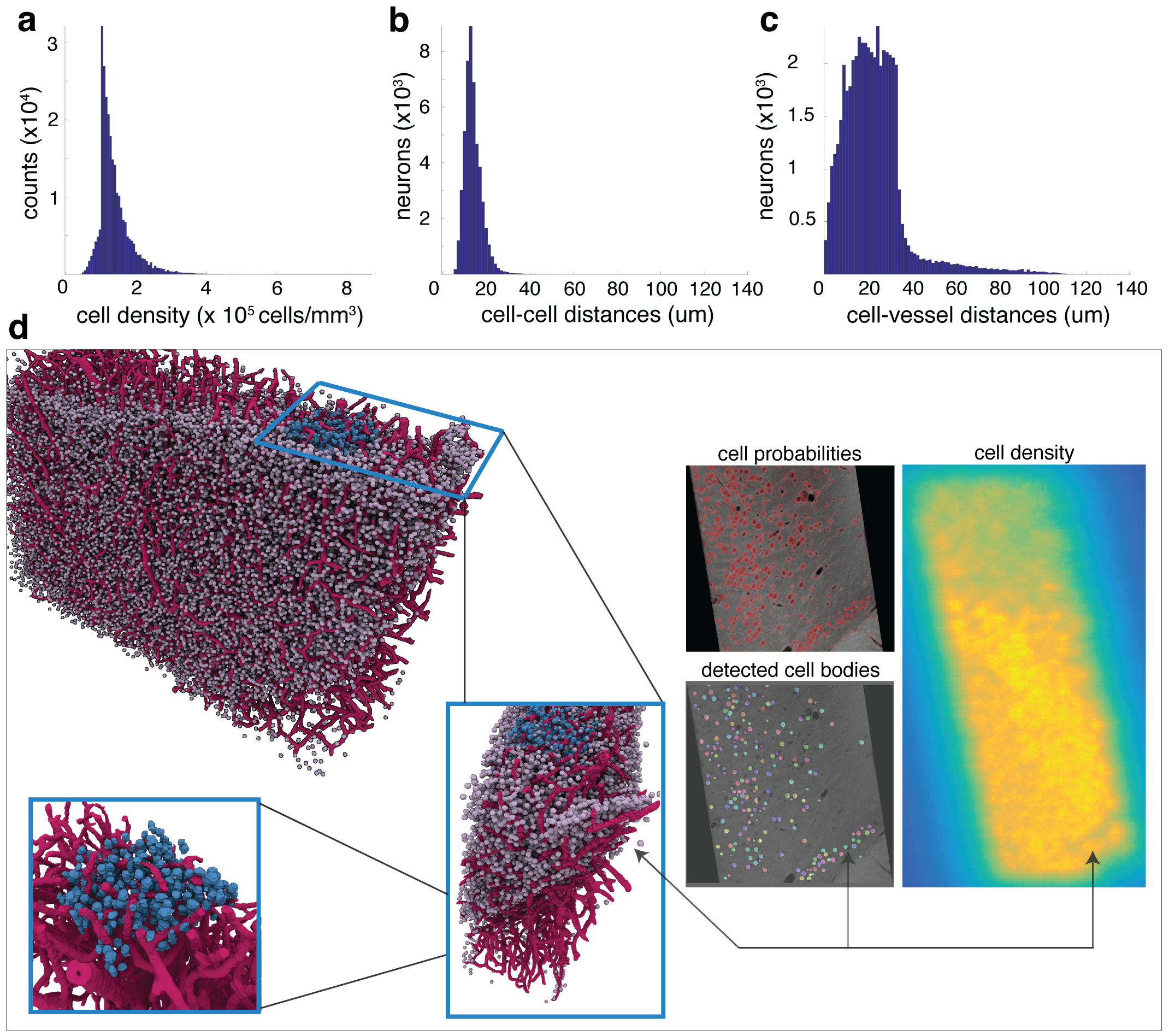}
\caption{ \small { \em Spatial statistics of X-ray volumes reveal layering and spatially-diverse distribution of cell bodies.} Along the top row, we display histograms of: (a) the estimates of the cell density over the extent of the entire sample of mouse cortex, (b) distances between the center of each cell and its nearest neighbor (cell-to-cell distances), and  (c) distances between the center of each cell and the closest vessel voxel (cell-to-vessel distances). In (d), we show multiple ways to visualize the data and confirm neuroantomical structure. We show a 3D rendering of the detected cells and vessels in the entire sample, with a manually labeled cube (V1) highlighted in blue. To confirm the 3D structure seen in these visualizations, on the right, we confirm the same 3D structure in the cell probability maps (red indicating high probability), detected cell maps (each detected cell displayed in a different color), and density estimates. This result provides further confirmation that the 3D structure of the sample is preserved in our density estimate. \label{fig:fig5}}
}
\end{figure}

The variation in training and test volume performance can be partially explained by fluctuations in the brightness, introduced during tomographic image reconstruction. To understand the connection between the fluctuations in contrast and difficulty of the cell detection problem, we computed the SNR across multiple cells within each of the labeled volumes. The mean and standard deviation of the signal-to-noise (SNR) between cells and their background in all three volumes was V1 = $(4.73, 0.69)$, V2 =$(4.59,1.49)$, and 
V3 = $(4.49,1.17)$. As expected, the precision and recall (for cell detection) seem to be correlated with the variance of the SNR in the volume (providing a measure fluctuations in contrast). In particular, we obtain the lowest precision and recall for V2, and indeed, this volume exhibits the highest variance in the contrast between cells and their background. Even in light of these fluctuations in brightness, our sensitivity analysis (Fig.\ \ref{fig:fig4}a) and results (Fig.\ \ref{fig:fig4}b) on training and test volumes suggest that X-BRAIN generalizes well across different regions of the volume. Furthermore, when we visually inspect our large-scale results (Fig. \ref{fig:fig4}c), we find a good correspondence between cells and vessels that are visible in slices and those detected by our algorithms. These results suggest that our algorithms are robust and can be applied at large scale.



We applied our pipeline to segment vessels and detect cells in a cubic mm sample ($2560 \times 2560 \times 1624$ voxels) of excised brain tissue collected from mouse somatosensory cortex (Fig.\ \ref{fig:fig4}). To apply X-BRAIN to large datasets, we created an analytics workflow that uses (but does not require) the LONI Pipeline environment\cite{rex2003loni} to automatically distribute jobs across a cluster environment.  Our workflow is parallelized by dividing our large dataset into small data blocks which can be processed independently, based upon a user-specified graphical (xml-based) description of the dependencies between various algorithms.  Running our analytics pipeline on a cubic mm sample took approximately six hours on a small 48-core cluster (see Methods). As a result, we detected $48$,$689$ cells over the extent of the analyzed sample ($\sim$0.42mm$^3$), which suggests a density of $1.18\times10^5$ cells per mm$^3$.

\subsection{Quantifying cellular and vascular densities and distances.} 

To compute the spatially-varying density of cells, we applied a robust non-parametric approach for density estimation. Adopting a non-parametric approach enables us to obtain an accurate estimate of the distribution without making any restrictive assumption on its form. In particular, we rely on the  popular $k$-nearest neighbors (kNN) density  estimation algorithm\cite{poczos2011estimation,loftsgaarden1965nonparametric}, which estimates a distribution using only distances between the samples (cells) and their $k^{\text{th}}$ nearest neighbor. When applied to the entire volume, we calculated an average density of $1.3 \times 10^5$ cells per mm$^3$ (Fig.\ \ref{fig:fig5}a). These results are comparable to other studies that estimate an average of $1.2\mbox{-}2.5 \times 10^5$ cells per mm$^3$ in mouse cortex\cite{tsai2009correlations}, both in terms of our average, and the spread in the distribution. These density estimates provide important information about the spatially-varying distribution of cells within the sample.



The location of cell bodies, relative to one another, and relative to the vasculature, is important for studying diseases that afflict the brain \cite{tsai2009correlations}. Thus we developed automated tools to compute distances between detected cell centers (cell-to-cell distances,, Fig.\ \ref{fig:fig5}b) and distances between each cell and the closest segmented vessel (cell-to-vessel distances, Fig.\ \ref{fig:fig5}c). Cell-to-vessel distances are spread between 10-40 $\mu$m, with very few cells exceeding this distance ($34.3\pm 533.4~\mum$). In contrast, the cell-to-cell distances appear to be much more concentrated, with a strong peak at $12.7 \mum$ and much smaller variance ($21.3\pm 43.1~\mum$). The distribution of distances between cells and vessels (Fig. \ref{fig:fig5}b) aligns with previous results\cite{tsai2009correlations,wu20143d} and confirms the accuracy of our approach for large-scale analysis. We further estimated that the fractional volume of vessels in the sample was $1.85\%$. This estimate is in agreement with previous studies\cite{heinzer2006hierarchical,tsai2009correlations,wu20143d}, which estimate the fractional density of vessels in the cortex to range from $0.97-3.64\%$. These results further confirm that our methods can be used to compute information about the relationship between cells and vasculature in the brain.

To complement our analysis tools, we developed methods to produce and visualize mesoscale maps, with the cellular density and vasculature as their output. These methods are integrated into the NeuroData framework and thus, after running a sample through our pipeline, users can download different descriptions of the neuroanatomy, either alone, or combined with the image data to help reveal relevant structures in the images. Using these multiple modes of visualization (Fig.~\ref{fig:fig5}d), we identified a 3D structure with extremely high cell density clustered at the bottom of the sample (Layer 6). We confirmed this structure in both 3D visualizations (left), in X-ray micrographs, the cell probability maps, and in our estimate of the cell density (right).  All of these representations provide information and descriptions of the data that can be used to further visualize and quantify its neuroanatomy. The combination of dense reconstructions of cells and blood vessels provide a unique approach for studying the joint distribution of brain cytoarchitecture and vasculature. 

\section{Discussion}
We have shown that $\mu$CT can be used to rapidly quantify mesoscale neuroanatomy in a millimeter scale sample without sectioning. Our results demonstrate how osmium stained and plastic embedded brains, in conjunction with a synchrotron X-ray source, produce sufficient contrast and resolution to automatically detect blood vessels and cell bodies. Our approach to automated anatomy is uniquely poised to provide detailed, large, mesoscale maps of the brain.

Our current approach does not yet provide enough resolution and contrast to resolve neural processes or reliably disambiguate cell types. In some cases  we can resolve large cellular process like apical dendrites, from which we can discern that the cell is indeed a neuron and potentially a pyramidal neuron. The resolution of $\mu$CT can be enhanced towards the nanometer scale, which would allow us to distinguish neuronal and non-neuronal types by shape\cite{peng2015bigneuron}. To complement cell type classification by shape, it is possible to develop genetic or immunohistochemical approaches for differential labeling of distinct neuronal and non-neuronal cell types for $\mu$CT.  Future applications of these techniques to enhance resolution and cell typing will enable a more detailed understanding of brain architecture through   $\mu$CT.

High-resolution approaches threaten to damage the specimen, as the X-ray dose increases quartically with the resolution\cite{howells_jesrp_2009}. For instance, beam damage can induce changes in sample geometry while the tomogram is being acquired, leading to reconstruction artifacts and the degradation of spatial resolution. However, the effect of radiation damage is greatly reduced in cold samples. As an example, the critical dose for mass loss in plastics (PMMA) is increased from 35 MGy at room temperature to 600 MGy when operating at a temperature of 113 K. X-rays have the potential for sub-$30$ nm resolution 3D imaging of frozen hydrated brain biopsies\cite{deng_pnas_2015} with no chemical modification or plastic embedding. Additional parameters of the imaging setup, including photon energy, coherence, and the optics can also be optimized to minimize damage while increasing resolution.

Our imaging data exhibits ringing artifacts that result from inhomogeneities in the signal source. The beamline used in our current study (2BM), uses a double multi-layer monochromator (DMM) to select a narrow bandwidth of the X-ray beam ($\Delta$E/E=$10^{-2}$). Multi-layers are known to introduce a significant amount of structures in the flat-field (the image without a sample). Such structures, when convolved with fluctuations of the source, generate imperfections in flat-field correction and produce ringing artifacts difficult to correct despite the use of ring artifact removal algorithms during image conversion (left, Fig.\ \ref{fig:fig1}c). Although our segmentation methods are relatively robust to these low-frequency artifacts, improvements to the tomography setup are important to improve data quality. Preliminary datasets collected at beamline 32-ID, a beamline that uses a short period undulator (U18) instead of a DMM, yield radiographs that are almost artifact free (see Supp.\ Fig.\ \ref{fig:suppfig}). Future improvements in source stability promise to yield tomographic reconstructions of considerably higher quality. 

Our current analysis pipeline is only designed to provide automatic segmentation of cells and vasculature. We currently do not address other aspects of the cytoarchitecture and vasculature important in neuroanatomy, such as cell shapes and neurites. Leveraging both computational methods and other histological preparations, we could develop more advanced approaches for distinguishing different cell types in X-ray tomograms. With significant further development and a massive-scale ground truthing effort, it should be possible to quantify the shape and morphology of neurons, as well as track neurites.

Limited ground truth data restricts in the complexity of methods that we can apply. With more training data, we can leverage more advanced nonlinear classification strategies, such as convolutional neural networks for segmentation and axon tracing. Such approaches have been shown to achieve state-of-the-art performance for identifying synapses and segmenting cell bodies in EM data\cite{grayroncal2015vesicle,turaga2010convolutional}. Finally, improvements in spatial resolution will help in the challenging problem of resolving adjacent neural structures as separate objects, leading to more straightforward and robust approaches for cell detection.

One of the advantages of our staining method is that we can apply $\mu$CT and automated EM imaging\cite{tapia2012high} to the same sample. Our results suggest that both techniques provide complementary images of the same sample without requiring modifications to existing EM techniques.  There are several potential advantages to combining these approaches: (i) providing large volume (but coarser resolution) inputs for EM reconstructions of synaptic connectivity, (ii) providing fiducials from inside the plastic block prior to deformation by cutting and thin section imaging for EM, and (iii) providing `pre-segmentation' of many cellular components to aid existing EM segmentation algorithms. Combining the advantages of $\mu$CT and EM promises to make both approaches stronger.

The high precision and recall of our algorithms suggest that our segmentation and cell detection methods can be used to reliably and quickly survey data volumes and identify cells and vessels in the sample. We can use these methods to build more systematic studies of regions of interest with EM, once the large-scale structure is identified using $\mu$CT. Information about where cells and vessels lie can be used as a prior in segmentation algorithms (in EM) and also to improve subsequent registration and alignment. As our pipeline for X-ray image analysis has been integrated in NeuroData, we can readily combine existing EM analysis pipelines in NeuroData with our methods to analyze the same dataset with $\mu$CT and EM. These results can be combined to create a multi-modal brain map that contains information about the cytoarchitectural and cerebrovascular properties of a sample, in addition to the fine-scale information about the processes and synapses afforded by EM.

Knowledge about the macro-scale organization of the brain, such as Brodmann maps\cite{zilles2010centenary}, have been based primarily on human anatomists working with thin, sparsely labeled slices of brains. However, with developments in large-scale connectomics with EM\cite{
helmstaedter2011high,roncal2015automated} and the techniques we present here for $\mu$CT, far larger and more comprehensive datasets become possible; datasets so large that no human could possibly digest them. The capabilities of synchotron source X-ray microscopy, combined with staining approaches for entire brain preparation \cite{mikula2015high}, offers the possibility of imaging entire brains at the mesoscale. With these capabilities, it should  become possible to obtain brain maps in a new, data-driven fashion, enabling the massive-scale quantification of a broad set of effects related to disease, development, and learning in the brain. 

\begin{methods}
We developed methods to image, segment, and analyze the neuroanatomical structure of brain volumes to quantify neuroanatomy with $\mu$CT. Our methods consist of three main parts: (1) sample preparation, (2) $\mu$CT and 3D image reconstruction, and (3) automatic segmentation and analysis of brain volumes.


\subsection{Sample preparation.}
To prepare the samples used in this paper, we used previous techniques for large volume EM\cite{Kasthuri2015}. Mice were anesthetized and transcardially perfused with aldehydes (2 percent PFA and 2 percent Glutaraldhyde), stained with heavy metals (osmium, uranium, and lead), dehydrated, and embedded in plastic (EPON). The main dataset analyzed in this paper is taken from mouse somatosensory cortex (S1). 

\subsection{Confirmation of cellular structures with subsequent EM.} After preparing the sample, we used synchrotron-based $\mu$CT to image 3D volumes of brain tissue at micron isotropic resolution. We subsequently ultra-thin sectioned the same tissue block with our approach to automated electron microscopy\cite{Kasthuri2015} and collected low-resolution EM micrographs ($\approx 100$ nm pixel resolution). In these low-resolution EM images, we identified the same pattern of cell bodies and vasculature that were found in the equivalent volume of X-ray data. Fine-resolution EM micrographs (3 nm pixel resolution) were then collected to identify synapses in the EM volumes.  Since these labeling approaches are species independent (i.e., they do not depend on transgenic strategies), we can apply this approach to human (and other) brain biopsies. 

\subsection{X-ray data collection and reconstruction.}
To collect the $\mu$CT datasets described here, we utilized the 2-BM beamline at the Advanced Photon Source (APS) with exposure times of $0.1$ second per projection and $3000$ projections. The 2-BM beamline uses a $10$ $\mum$ thick LuAG:Ce scintillator to convert propagation-enhanced X-ray wave into visible light, which a microscope objective magnifies onto a visible light-scientific CMOS camera (pco.edge $5.5$ camera with $2560 \times 2560$ pixels). When using a 10X objective, this yields a projection with a pixel size of $0.65$ $\mum$.  We utilized propagation-based phase contrast X-ray imaging to obtain high-contrast tomograms of millimeter-sized regions of plastic embedded and metal-stained mouse brain. Imaging a $1$ ${\rm mm}^3$ volume at 1 micron isotropic takes approximately six minutes and requires no volume alignment or registration process post-acquisition. The X-ray energy bandwidth was about $300$ eV, which means that the data are largely free of the ``beam hardening'' effect that otherwise complicates medical imaging using laboratory X-ray sources.\footnote{X-ray spectrum changes with depth in tissue as the lower energy X-rays are absorbed, leading to changes in the Fresnel fringes used to obtain phase contrast.}  We are thus able to obtain data around $130$ times faster than with laboratory sources, and with higher image quality.

\subsection{Reconstruction of 3D volumes.} 
Datasets were collected in Hierarchical Data Format (HDF) files using the Data Exchange schema developed for synchrotron data\cite{de2014scientific}. Data processing and image reconstructions were performed using the TomoPy toolbox, an open-source Python package, developed at the Advanced Photon Source (APS) for tomographic data analysis\cite{gursoy2014tomopy}. We first normalized the projection images with the incident X-ray measurements to suppress artifacts originating from imperfections in the detection process. A wavelet-Fourier filter\cite{mvrnch2009stripe} is used to further suppress these artifacts with ten wavelet levels and an offset suppression value of two. We used a Paganin-type single-step phase retrieval algorithm to retrieve the phase of the transmitted X-ray signal\cite{paganin2002simultaneous}. The location of the rotation center was estimated either automatically --- using an optimization approach minimizing the entropy in reconstructions\cite{donath2006automated}, or manually --- if signal-to-noise levels are high. The tomographic reconstructions were performed using GridRec algorithm\cite{dowd1999developments}, which is a fast implementation of the conventional filtered-back-projection method\cite{avinash1988principles}.

\subsection{Preprocessing of image stacks.} Each image reconstructed in TomoPy is $2560 \times 2560$ pixels ($0.65~ \mum$ isotropic) and is initially stored with 32-bit float precision. We utilized the multiple image processor tool in Fiji (ImageJ)\cite{schindelin2012fiji} to convert the bit depth of each $\mu$CT image to 8-bit images. By computing the average number of bits of information in each pixel of the original image, we confirmed that an 8-bit depth was sufficient to capture the information in the $\mu$CT stack. Visual inspection also confirmed this choice of bit depth, with no visible loss of data quality due to quantization. Following this, we applied an automatic contrast enhancement filter to each image in the stack in Fiji.

\subsection{Semi-supervised masking protocol.} To reduce the size of the data and the complexity of subsequent processing steps, we applied a semi-supervised masking algorithm to identify regions in each image where the biological specimen is present (versus background pixels). We developed a supervised method, which requires a user to draw a mask between a small number of dataset images and then finds a smooth interpolation of the mask between these labeled frames. This toolbox utilizes the roipoly tool in Matlab, which provides a GUI for drawing a polygon around the brain tissue in each image. 
The semi-automated masking step is a small part of the overall workflow and is only performed once per dataset.  After reducing the bit depth and masking the data, the dataset is reduced from 95 GB to only 10 GB. 

\subsection{Volume of the analyzed sample.}
The image volume that we analyze in this paper is of size $1400\times2480\times1547$ voxels, which corresponds to a volume of size $910 \times 1612 \times 1005$ microns ($1.474$ cubic millimeters). As the sample is rotated within the field of view (sample plane), we compute that the number of unmasked voxels represents a volume of approximately $0.41$ cubic mm.

\subsection{Storage of $\mu$CT data and annotations in NeuroData.}  We uploaded the raw and masked images into an open-access platform for neuroimaging datasets called NeuroData(\url{www.neurodata.io}). Additionally, we also store the annotations and segmentations resulting from our analysis in the NeuroData spatial database to facilitate rapid access, dissemination, and analysis of the data. This framework allows for researchers to freely download arbitrary volumes of raw data, manual labels, or automated annotations for algorithm development or analysis.  Users may also query the metadata of detected cells within a volume, which enables rapid knowledge extraction from the X-ray datasets and statistical analyses at scale.

\subsection{Data accessibility and reproducibility.}  Our methodology, end-to-end pipeline, algorithms, data, and data derivatives are all open source and available for others to reproduce and leverage for further scientific discovery.  As described above, we facilitate this open access via integration with the NeuroData ecosystem, which provides tools and infrastructure to store, visualize, parse, and analyze big neuroscience data\cite{burns2013open}.  Both the raw data and its derivatives are freely available on NeuroData for download and visualization (Fig.\ \ref{fig:suppfig2} in Supp.\ Materials).

\subsection{Evaluation metrics.} To compute human-to-human agreement and evaluate the performance of our methods, we developed tools to compare segmentations at both the pixel and object level. Detected pixels/objects that do not appear in the manual segmentation are counted as false positives, and manually identified pixels/objects not found by the automatic segmentation algorithm result in  false negatives (misses). In all of our evaluations, we compute the precision ($p$), recall ($r$), and f$_\beta$ score 
$$f_{\beta} = (1 + \beta^2)\frac{pr}{\beta^2p + r},$$
where we set $\beta = \{1,2\}$. When evaluating the performance of our methods for detecting cells (object-level errors), we compute matches between two sets of centroids by identifying cell pairs in different segmentations that are nearest neighbors.  If the matching centroids are within a fixed distance ($10~\mum$) from one another, we label them a match and remove both cells from the dataset to avoid duplicate assignments. The matching process iterates until all possible matches are found, and precision and recall metrics are computed. For cell detection, we compute the $f_1$ score as it places equal weight on precision and recall. However, in the case of the pixel-level segmentation of vessels, we observe that optimizing the $f_2$ score produces more accurate results (confirmed by visual inspection). 

\subsection{Manual labeling and human-to-human agreement.}

In order to obtain ground truth datasets to quantify the performance of our algorithms and to assess human-to-human agreement, we used a total of four trained annotators (A0, A1, A2, A3) and five novices to label different sub-volumes (V0, V1, V2, V3) of our image dataset using ITK-Snap\cite{py06nimg}.

Two of the trained experts (A0, A1) and the five novices labeled cells and vessels in V1, a $195\times 195\times 65$ micron cube of data ($300\times 300\times 100$ voxels). Annotator A0 was instructed to produce a {\em saturated reconstruction}, where all cells and vessels were fully labeled. A1 produced a saturated segmentation of a sub-volume of V1, which we denote as V0. To then compute human-to-human agreement across annotators, we computed the voxel-wise precision and recall between V0-A0 (ground truth) and V0-A1, which we compute to be $(p,r) = (0.93,0.58)$ for cell bodies and $(p,r) = (0.99,0.29)$ for vessels. While precision is high in both cases, the recall is much lower. This is due to the fact that A1 produces an underestimate of A0's labels; we tested this by dilating A1's labels until we maximized the $f_1$ score between both annotations. In this case, we obtain a  precision $(p,r) = (0.84,0.76)$ for cell bodies and $(p,r) = (0.85,0.73)$ for vessels.

We then computed the agreement between these annotators in detecting cell centroids. We first cleaned each segmentation to ensure all cells are disconnected from one another. We then applied a connected component algorithm and found the center of mass of each component to estimate the centroid of each cell. We matched centroids across the two annotations and computed object-level precision and recall. When ignoring cells along the boundaries of the volume, there are no cells identified by A1 that are not identified by A0 and only one cell identified by A0 that was not identified by A1. Thus, human-to-human agreement is nearly perfect when asked to identify cell centers.

In addition to computing human agreement, we also acquired additional volumes for testing our algorithms. For these purposes, we had another expert annotator (A2) densely label the cells and partially label the vessels in a training volume (V2). Annotator A1 then edited all cells in  this volume, which we denoted as V2-A12. Finally, to test our methods, we had an external party randomly select a sub-volume to be used as a hold-out test volume at a location unknown to the authors of this manuscript. Annotator A0 and A3 then iteratively refined a common estimate of the cell centroids in volume (V3); this annotation is referred to as V3-A03. V3 is used as a hold-out test set to evaluate the accuracy of our cell detection method. 

To quantify the time required to label the centroids of cell bodies, we recruited five subjects with no previous experience to label the centers of cell bodies in 3D. Each subject was instructed to label as many cells as possible in thirty minutes. The average number of cells that these subjects labeled was $51.2$ and the median was $62$. These results suggests that a novice can label the centroids of around $100$ cells in one hour. In practice, we find that it takes experts around 5 hours to reliably label all cell centers in a $(100~\mum)^3$ volume. From estimates of the cell density in mouse cortex, we expect around 120,000 cells per cubic millimeter; therefore, to manually annotate all cells in a cubic mm would require a projected 1200 person-hours or 50 days working 24 hours per day.

\subsection{Computing the signal-to-background (SNR).} To estimate the intrinsic difficulty of separating cells from their background, we calculated the ratio of the intensity between cells and their exteriors. To do this, we sampled 10 cells every 25 slices ($15.6~\mum$) in each of the three manually annotated volumes (V1, V2, V3) using ITK Snap. We placed a small circular marker within the cell's boundary and a marker outside of the cell in a location where the cell's boundary is clearly resolved. This generated 30 samples in both V1 and V2 and 89 samples in V3, of the brightness inside (signal) and outside (noise). We then computed the signal-to-noise ratio (SNR) for the $i^{\rm th}$ cell as follows:
$$ {SNR} = 20 \log_{10} \bigg( 
\frac{s_i}{n_i}
\bigg), $$
where $s_i$ (signal) and $n_i$ (noise) contains the mean value of the labeled pixels within and outside of the $i^{\rm th}$ labeled cell, respectively. The mean and standard deviation of the SNR (dB) across each subvolume is:
V1 = $(4.73,0.69)$, V2 = $(4.59,1.49)$, V3 = $(4.49,1.13)$. Thus, we observe the largest variance in SNR in V2 and the lowest average SNR in V3. The training volume V1 appears to have the highest mean and lowest variance out of the three volumes. Our estimates of SNR appear to be predictive of the difficulty of the segmentation task, and thus are correlated with the accuracy of our segmentation results on the different volumes.

\subsection{X-BRAIN: Methods for segmenting and analyzing X-ray image volumes.}
To facilitate biological interpretation and knowledge extraction from $\mu$CT datasets, we developed a suite of computer vision and image processing methods to segment and analyze mesoscale data. We refer to this set of tools as X-BRAIN (X-ray Brain Reconstruction, Analytics, and Inference for Neuroscience).
We now provide an overview of the modules and tools provided in X-BRAIN. Following this, we provide additional details about this toolkit.
\vspace{-2mm}
\begin{itemize}
\item{\bf Data download:} Using Neurodata web-services, we provide scripts for accessing subvolumes of the raw and masked data. 
\item {\bf Segmentation:} Using the ilastik\cite{ilastik} classification tool, we produce a three-class probability map which encodes the probability of each voxel belonging to class `cell,' `blood vessel,' or `other' (lies in  background). We use these probability maps to obtain a segmentation of vessels and cells in the volume.
\item {\bf Estimating the size and location of cell bodies:} To estimate the position of cells in the volume, we develop an iterative approach for cell detection that applies a fast-3D convolution method to detect cell bodies based on the ilastik probability maps. We then estimate the size of each cell based upon the detected centers.
\item {\bf Methods for computing spatial statistics:} We apply a robust density estimation technique to detected centroids, which provides a non-parametric estimate of the underlying density of cells in the sample. Additionally, we provide methods to compute the distance between detected cells and vessels in addition to spatially-varying vessel density measures.
\item {\bf Data upload:} The raw and masked image datasets, cell and vessel probabilities and 3D segmentations are uploaded to NeuroData spatial databases to allow for later queries and analysis. 
\end{itemize}

\noindent{\em (1) Computing probability maps with ilastik.}
The first step of our segmentation pipeline is to perform pixel-level classification on the X-ray images (2D), which provides the posterior probablities that each voxel is either a cell, vessel, or lies in the background (other). We applied a tool called ilastik which trains a random forest classifier from sparse (manual) annotations of class labels in the data.  Ilastik provides an interactive method to compute and examine feature channels; we selected a variety of patch-based texture features at different scales for this analysis.  

\noindent{\em (2) Vessel segmentation.}
After computing the vessel probability map with ilastik, we threshold the probability map (see Fig.\ \ref{fig:fig4}b to assess the sensitivity of our algorithm to different choices of thresholds), dilate the resulting binary thresholded output, and then remove spurrious connected components based on a minimum size threshold. After applying these simple morphological filtering operations, we find that the resulting segmentation has a higher agreement with the (manually segmented) ground truth than the labels produced by a second manual annotator.

\noindent{\em (3) Greedy sphere finding approach for cell detection.}
While ilastik provides a good starting point for identifying cell body locations, individual cells and vessels are often hard to distinguish by simply thresholding the probability map. To separate these components into their constituent parts (cells and vessels), we developed a greedy approach which is similar in spirit to matching pursuit algorithms for sparse signal recovery\cite{davis1997adaptive}. The main idea behind our approach for cell finding is to iteratively refine our estimate of the cell position and then ``remove'' this cell from the data. We do this by first creating a spherical template with a diameter roughly equal to that of the cell; the exact choice of parameter was learned through a hyperparameter search (see below).  We apply a 3D-FFT to convolve the spherical template with the cell probability map produced by ilastik. This produces a ``sphere map'' which gives us high responses in regions that are likely to contain cell bodies. At each step of our algorithm, we select the global maxima of the sphere map to be the centroid of the next detected cell. After finding this cell, we then zero out the probability map in this region so that we cannot select a candidate cell in this same location again, and repeat this matching procedure until convergence. We define convergence as the point at which the correlation between the probability map and our template drops below a user-specified threshold or reaches the maximum number of iterations.

\noindent{\em (4) Hyperparameter searches.} We developed a tool to run hyperparameter searches over our methods to maximize performance on the ground truth volume V1.  After exploring the parameter space, we ran a grid search over the most critical parameters (cell size, dilation, and threshold cutoff) to find a stable, optimal point. We selected the parameters (cell size: 18, dilation: 8, threshold 0.47) that maximized $f_1$, the harmonic mean between precision and recall. Because voxels on the edge of volumes have inherent ambiguity for both human and machine annotators, we choose to disregard objects at the edge (of both detected and truth volumes) when computing precision and recall scores throughout this manuscript to ensure the most representative result.

\newcommand{\x}{{\bf x}}
\newcommand{\A}{{\bf A}}
\renewcommand{\a}{{\bf a}}
\renewcommand{\v}{{\bf v}}
\newcommand{\V}{{\bf V}}
\newcommand{\R}{\mathbb{R}}

\noindent{\em (5) Non-parametric density estimation.}
To compute the density of detected cells within a volume, we applied a $k$-nearest neighbor (kNN) density  estimation algorithm\cite{poczos2011estimation,loftsgaarden1965nonparametric}, which estimates the density using only distances between the samples (cells) and their $k^{\text{th}}$ nearest neighbor. More concretely, we define the distance between a centroid vector $\x \in \R^3$ and a matrix $\A$ as $$\rho_k(\x , \A ) = \| \x - \a_k \|^2_2,$$  where $\a_k$ is the $k^{\rm th}$ nearest neighbor to $\x$ contained in the columns of $\A$. The value of the empirical distribution $p$ at $\v = (x,y,z)$ is then estimated using the following consistent estimator\cite{poczos2011estimation}:
\begin{equation*}
p( \v) \propto \frac{k}{ N \rho_k( \v , \V )},
\end{equation*}
and $\V$ contains the centroids of the rest of the detected cells in the sample. We compute this quantity over a 3D grid, where the volume of each bin in the sample grid is ${\rm Vol} = (8.44 \mum)^3$. We selected this bin size to ensure that detected cells will lie in roughly a single grid point. This choice was further confirmed by visually inspecting the resulting density estimates. After computing the density for each 3D bin in our selected grid, we normalize to obtain a proper probability density function. Finally, we compute an estimate of the number of cells per cubic mm as, $p_d(\v) = (p(\v)N/\mbox{Vol})\times 10^9$. The intuition behind this approach is that in regions where we have higher density of samples, the quantity $\rho_k( \v_i, \V)$ will be very small and thus, the probability of generating a sample at this location is large. In practice, we set $k = \sqrt{N}$ which guarantees that the estimates of $p$ will asymptotically converge to the exact point estimates of the distribution since $\rho_k$ converges to $0$ as $N\to+\infty$\cite{poczos2011estimation}.

\subsection{Details of experiments on large-scale datasets.}
After validating and benchmarking our algorithms, we scaled our processing to the entire dataset of interest (x voxels: 610-2010, y: 1-2480, z: 390-2014, resolution: 0), using the LONI processing environment\cite{rex2003loni}.  Leveraging LONI allows us to quickly build interfaces to algorithms written by different research groups and in different languages to assemble a cohesive pipeline.  These algorithms have well-defined interfaces and can also be repackaged for use in a different meta-scheduler environment.  When running at scale, we first chunked our data into small cuboids meeting our computational constraints, and then ran each block through the pipeline.  NeuroData was used to get and store data; image data was requested for each computed block and the results were written to a spatially co-registered annotation channel\cite{burns2013open}.  Each block was retrieved with sufficient padding to provide edge context; we  processed these blocks in a parallel fashion and uploaded the resulting detections to NeuroData.  We have also implemented an alternative merging strategy to account for cells near boundaries.  Briefly, we eliminate putative detections touching an edge or that overlap an object already present in the database to further reduce edge effects.

\end{methods}

\begin{addendum}
\item[Acknowledgements.] This research used resources of the U.S. Department of Energy (DOE) Office of Science User Facilities operated for the DOE Office of Science by Argonne National Laboratory under Contract No. DE-AC02-06CH11357. Support was provided by: NIH U01MH109100 (ELD, HLF, DG, XX, CJ, NK, and KK), the IARPA MICRONS project (NK), an educational fellowship from the Johns Hopkins University Applied Physics Laboratory (WGR), the Defense Advanced Research Projects Agency (DARPA) SIMPLEX program through SPAWAR contract N66001-15-C-4041, and DARPA GRAPHS N66001-14-1-4028. The sample showed in the Supp.\ Materials was graciously provided by Haruo Mizutani. The authors would like to thank Jordan Matelsky from JHU and NeuroData for all of his help in creating 3D visualizations of the data in Blender. We would also like to thank Norman Clark and Olivia Grebski for their assistance in annotating some of the training volumes. A big thanks to Francesco De Carlo (Argonne) and Pavan Ramkumar (Northwestern) for useful discussions regarding this work.

 \item[Competing Interests] The authors declare that they have no
competing financial interests.

 \item[Correspondence] Correspondence and requests for materials
should be addressed to E.L.D. and N.K. \\~(email: edyer@northwestern.edu, bobbykasthuri@anl.gov).
\end{addendum}

\section*{References}
\vspace{8mm}
\bibliography{xbrain-manuscript-v0}

\begin{thebibliography}{10}
\expandafter\ifx\csname url\endcsname\relax
  \def\url#1{\texttt{#1}}\fi
\expandafter\ifx\csname urlprefix\endcsname\relax\def\urlprefix{URL }\fi
\providecommand{\bibinfo}[2]{#2}
\providecommand{\eprint}[2][]{\url{#2}}

\bibitem{lichtman2011big}
\bibinfo{author}{Lichtman, J.~W.} \& \bibinfo{author}{Denk, W.}
\newblock \bibinfo{title}{The big and the small: challenges of imaging the
  brain's circuits}.
\newblock \emph{\bibinfo{journal}{Science}} \textbf{\bibinfo{volume}{334}},
  \bibinfo{pages}{618--623} (\bibinfo{year}{2011}).

\bibitem{economo2016platform}
\bibinfo{author}{Economo, M.~N.} \emph{et~al.}
\newblock \bibinfo{title}{A platform for brain-wide imaging and reconstruction
  of individual neurons}.
\newblock \emph{\bibinfo{journal}{eLife}} \textbf{\bibinfo{volume}{5}},
  \bibinfo{pages}{e10566} (\bibinfo{year}{2016}).

\bibitem{amunts2013bigbrain}
\bibinfo{author}{Amunts, K.} \emph{et~al.}
\newblock \bibinfo{title}{Bigbrain: an ultrahigh-resolution {3D} human brain
  model}.
\newblock \emph{\bibinfo{journal}{Science}} \textbf{\bibinfo{volume}{340}},
  \bibinfo{pages}{1472--1475} (\bibinfo{year}{2013}).

\bibitem{eberle2015high}
\bibinfo{author}{Eberle, A.} \emph{et~al.}
\newblock \bibinfo{title}{High-resolution, high-throughput imaging with a
  multibeam scanning electron microscope}.
\newblock \emph{\bibinfo{journal}{Journal of Microscopy}}
  \textbf{\bibinfo{volume}{259}}, \bibinfo{pages}{114--120}
  (\bibinfo{year}{2015}).

\bibitem{chung2013clarity}
\bibinfo{author}{Chung, K.} \& \bibinfo{author}{Deisseroth, K.}
\newblock \bibinfo{title}{Clarity for mapping the nervous system}.
\newblock \emph{\bibinfo{journal}{Nature {M}ethods}}
  \textbf{\bibinfo{volume}{10}}, \bibinfo{pages}{508--513}
  (\bibinfo{year}{2013}).

\bibitem{chen2015expansion}
\bibinfo{author}{Chen, F.}, \bibinfo{author}{Tillberg, P.~W.} \&
  \bibinfo{author}{Boyden, E.~S.}
\newblock \bibinfo{title}{Expansion microscopy}.
\newblock \emph{\bibinfo{journal}{Science}} \textbf{\bibinfo{volume}{347}},
  \bibinfo{pages}{543--548} (\bibinfo{year}{2015}).

\bibitem{bushong2015x}
\bibinfo{author}{Bushong, E.~A.} \emph{et~al.}
\newblock \bibinfo{title}{{X}-ray microscopy as an approach to increasing
  accuracy and efficiency of serial block-face imaging for correlated light and
  electron microscopy of biological specimens}.
\newblock \emph{\bibinfo{journal}{Microscopy and Microanalysis}}
  \textbf{\bibinfo{volume}{21}}, \bibinfo{pages}{231--238}
  (\bibinfo{year}{2015}).

\bibitem{mikula2015high}
\bibinfo{author}{Mikula, S.} \& \bibinfo{author}{Denk, W.}
\newblock \bibinfo{title}{High-resolution whole-brain staining for electron
  microscopic circuit reconstruction}.
\newblock \emph{\bibinfo{journal}{Nature {M}ethods}}
  \textbf{\bibinfo{volume}{12}}, \bibinfo{pages}{541--546}
  (\bibinfo{year}{2015}).

\bibitem{arillo_sysent_2015}
\bibinfo{author}{Arillo, A.} \emph{et~al.}
\newblock \bibinfo{title}{Long-proboscid brachyceran flies in {Cretaceous}
  amber ({D}iptera: {S}tratiomyomorpha: {Z}hangsolvidae)}.
\newblock \emph{\bibinfo{journal}{Systematic Entomology}}
  \textbf{\bibinfo{volume}{40}}, \bibinfo{pages}{242--267}
  (\bibinfo{year}{2015}).

\bibitem{mizutani2009microtomographic}
\bibinfo{author}{Mizutani, R.} \emph{et~al.}
\newblock \bibinfo{title}{Microtomographic analysis of neuronal circuits of
  human brain}.
\newblock \emph{\bibinfo{journal}{Cerebral Cortex}}  (\bibinfo{year}{2009}).

\bibitem{mizutani2010unveiling}
\bibinfo{author}{Mizutani, R.} \emph{et~al.}
\newblock \bibinfo{title}{Unveiling 3{D} biological structures by {X}-ray
  microtomography}.
\newblock \emph{\bibinfo{journal}{Microscopy: Science, Technology, Applications
  and Education}} \bibinfo{pages}{379--386} (\bibinfo{year}{2010}).

\bibitem{mizutani2012x}
\bibinfo{author}{Mizutani, R.} \& \bibinfo{author}{Suzuki, Y.}
\newblock \bibinfo{title}{{X}-ray microtomography in biology}.
\newblock \emph{\bibinfo{journal}{Micron}} \textbf{\bibinfo{volume}{43}},
  \bibinfo{pages}{104--115} (\bibinfo{year}{2012}).

\bibitem{de2014scientific}
\bibinfo{author}{De~Carlo, F.} \emph{et~al.}
\newblock \bibinfo{title}{Scientific data exchange: a schema for {HDF}5-based
  storage of raw and analyzed data}.
\newblock \emph{\bibinfo{journal}{Journal of Synchrotron Radiation}}
  \textbf{\bibinfo{volume}{21}}, \bibinfo{pages}{1224--1230}
  (\bibinfo{year}{2014}).

\bibitem{tapia2012high}
\bibinfo{author}{Tapia, J.~C.} \emph{et~al.}
\newblock \bibinfo{title}{High-contrast en bloc staining of neuronal tissue for
  field emission scanning electron microscopy}.
\newblock \emph{\bibinfo{journal}{Nature {P}rotocols}}
  \textbf{\bibinfo{volume}{7}}, \bibinfo{pages}{193--206}
  (\bibinfo{year}{2012}).

\bibitem{paganin2002simultaneous}
\bibinfo{author}{Paganin, D.}, \bibinfo{author}{Mayo, S.},
  \bibinfo{author}{Gureyev, T.~E.}, \bibinfo{author}{Miller, P.~R.} \&
  \bibinfo{author}{Wilkins, S.~W.}
\newblock \bibinfo{title}{Simultaneous phase and amplitude extraction from a
  single defocused image of a homogeneous object}.
\newblock \emph{\bibinfo{journal}{Journal of Microscopy}}
  \textbf{\bibinfo{volume}{206}}, \bibinfo{pages}{33--40}
  (\bibinfo{year}{2002}).

\bibitem{gursoy2014tomopy}
\bibinfo{author}{G{\"u}rsoy, D.}, \bibinfo{author}{De~Carlo, F.},
  \bibinfo{author}{Xiao, X.} \& \bibinfo{author}{Jacobsen, C.}
\newblock \bibinfo{title}{Tomopy: a framework for the analysis of synchrotron
  tomographic data}.
\newblock \emph{\bibinfo{journal}{Journal of Synchrotron Radiation}}
  \textbf{\bibinfo{volume}{21}}, \bibinfo{pages}{1188--1193}
  (\bibinfo{year}{2014}).

\bibitem{zhang_jvstb_1995}
\bibinfo{author}{Zhang, X.}, \bibinfo{author}{Jacobsen, C.},
  \bibinfo{author}{Lindaas, S.} \& \bibinfo{author}{Williams, S.}
\newblock \bibinfo{title}{Exposure strategies for polymethyl methacrylate from
  {\emph{in situ}} {X}-ray absorption near edge structure spectroscopy}.
\newblock \emph{\bibinfo{journal}{Journal of Vacuum Science and Technology B}}
  \textbf{\bibinfo{volume}{13}}, \bibinfo{pages}{1477--1483}
  (\bibinfo{year}{1995}).

\bibitem{williams_jm_1993}
\bibinfo{author}{Williams, S.} \emph{et~al.}
\newblock \bibinfo{title}{{Measurements of wet metaphase chromosomes in the
  scanning transmission {X}-ray microscope}}.
\newblock \emph{\bibinfo{journal}{Journal of Microscopy}}
  \textbf{\bibinfo{volume}{170}}, \bibinfo{pages}{155--165}
  (\bibinfo{year}{1993}).

\bibitem{rose_jsmpe_1946}
\bibinfo{author}{Rose, A.}
\newblock \bibinfo{title}{A unified approach to the performance of photographic
  film, television pickup tubes, and the human eye}.
\newblock \emph{\bibinfo{journal}{Journal of the Society of Motion Picture
  Engineers}} \textbf{\bibinfo{volume}{47}}, \bibinfo{pages}{273--294}
  (\bibinfo{year}{1946}).

\bibitem{py06nimg}
\bibinfo{author}{Yushkevich, P.~A.} \emph{et~al.}
\newblock \bibinfo{title}{User-guided {3D} active contour segmentation of
  anatomical structures: Significantly improved efficiency and reliability}.
\newblock \emph{\bibinfo{journal}{Neuroimage}} \textbf{\bibinfo{volume}{31}},
  \bibinfo{pages}{1116--1128} (\bibinfo{year}{2006}).

\bibitem{ilastik}
\bibinfo{author}{Sommer, C.}, \bibinfo{author}{Straehle, C.},
  \bibinfo{author}{Koethe, U.} \& \bibinfo{author}{Hamprecht, F.~A.}
\newblock \bibinfo{title}{ilastik: Interactive learning and segmentation
  toolkit}.
\newblock In \emph{\bibinfo{booktitle}{IEEE International Symposium on
  Biomedical Imaging}} (\bibinfo{year}{2011}).

\bibitem{rex2003loni}
\bibinfo{author}{Rex, D.~E.}, \bibinfo{author}{Ma, J.~Q.} \&
  \bibinfo{author}{Toga, A.~W.}
\newblock \bibinfo{title}{The loni pipeline processing environment}.
\newblock \emph{\bibinfo{journal}{Neuroimage}} \textbf{\bibinfo{volume}{19}},
  \bibinfo{pages}{1033--1048} (\bibinfo{year}{2003}).

\bibitem{poczos2011estimation}
\bibinfo{author}{P{\'o}czos, B.} \& \bibinfo{author}{Schneider, J.~G.}
\newblock \bibinfo{title}{On the estimation of alpha-divergences}.
\newblock In \emph{\bibinfo{booktitle}{Int. Conf. on Artificial Intelligence
  and Statistics}}, \bibinfo{pages}{609--617} (\bibinfo{year}{2011}).

\bibitem{loftsgaarden1965nonparametric}
\bibinfo{author}{Loftsgaarden, D.~O.}, \bibinfo{author}{Quesenberry, C.~P.}
  \emph{et~al.}
\newblock \bibinfo{title}{A nonparametric estimate of a multivariate density
  function}.
\newblock \emph{\bibinfo{journal}{Ann. Math. Stat.}}
  \textbf{\bibinfo{volume}{36}}, \bibinfo{pages}{1049--1051}
  (\bibinfo{year}{1965}).

\bibitem{tsai2009correlations}
\bibinfo{author}{Tsai, P.~S.} \emph{et~al.}
\newblock \bibinfo{title}{Correlations of neuronal and microvascular densities
  in murine cortex revealed by direct counting and colocalization of nuclei and
  vessels}.
\newblock \emph{\bibinfo{journal}{The Journal of Neuroscience}}
  \textbf{\bibinfo{volume}{29}}, \bibinfo{pages}{14553--14570}
  (\bibinfo{year}{2009}).

\bibitem{wu20143d}
\bibinfo{author}{Wu, J.} \emph{et~al.}
\newblock \bibinfo{title}{3d brain{CV}: simultaneous visualization and analysis
  of cells and capillaries in a whole mouse brain with one-micron voxel
  resolution}.
\newblock \emph{\bibinfo{journal}{Neuroimage}} \textbf{\bibinfo{volume}{87}},
  \bibinfo{pages}{199--208} (\bibinfo{year}{2014}).

\bibitem{heinzer2006hierarchical}
\bibinfo{author}{Heinzer, S.} \emph{et~al.}
\newblock \bibinfo{title}{Hierarchical microimaging for multiscale analysis of
  large vascular networks}.
\newblock \emph{\bibinfo{journal}{Neuroimage}} \textbf{\bibinfo{volume}{32}},
  \bibinfo{pages}{626--636} (\bibinfo{year}{2006}).

\bibitem{peng2015bigneuron}
\bibinfo{author}{Peng, H.} \emph{et~al.}
\newblock \bibinfo{title}{Bigneuron: large-scale 3{D} neuron reconstruction
  from optical microscopy images}.
\newblock \emph{\bibinfo{journal}{Neuron}} \textbf{\bibinfo{volume}{87}},
  \bibinfo{pages}{252--256} (\bibinfo{year}{2015}).

\bibitem{howells_jesrp_2009}
\bibinfo{author}{Howells, M.~R.} \emph{et~al.}
\newblock \bibinfo{title}{An assessment of the resolution limitation due to
  radiation-damage in {X}-ray diffraction microscopy}.
\newblock \emph{\bibinfo{journal}{Journal of Electron Spectroscopy and Related
  Phenomena}} \textbf{\bibinfo{volume}{170}}, \bibinfo{pages}{4--12}
  (\bibinfo{year}{2009}).

\bibitem{deng_pnas_2015}
\bibinfo{author}{Deng, J.} \emph{et~al.}
\newblock \bibinfo{title}{Simultaneous cryo {X}-ray ptychographic and
  fluorescence microscopy of green algae}.
\newblock \emph{\bibinfo{journal}{Proceedings of the National Academy of
  Sciences}} \textbf{\bibinfo{volume}{112}}, \bibinfo{pages}{2314--2319}
  (\bibinfo{year}{2015}).

\bibitem{grayroncal2015vesicle}
\bibinfo{author}{{Gray Roncal}, W.} \emph{et~al.}
\newblock \bibinfo{title}{{VESICLE : Volumetric Evaluation of Synaptic
  Inferfaces using Computer vision at Large Scale}}.
\newblock In \emph{\bibinfo{booktitle}{26th British Machine Vision Conference
  (BMVC)}}, \bibinfo{pages}{1--9} (\bibinfo{year}{2015}).

\bibitem{turaga2010convolutional}
\bibinfo{author}{Turaga, S.~C.} \emph{et~al.}
\newblock \bibinfo{title}{Convolutional networks can learn to generate affinity
  graphs for image segmentation}.
\newblock \emph{\bibinfo{journal}{Neural computation}}
  \textbf{\bibinfo{volume}{22}}, \bibinfo{pages}{511--538}
  (\bibinfo{year}{2010}).

\bibitem{zilles2010centenary}
\bibinfo{author}{Zilles, K.} \& \bibinfo{author}{Amunts, K.}
\newblock \bibinfo{title}{Centenary of brodmann's map—conception and fate}.
\newblock \emph{\bibinfo{journal}{Nature Reviews Neuroscience}}
  \textbf{\bibinfo{volume}{11}}, \bibinfo{pages}{139--145}
  (\bibinfo{year}{2010}).

\bibitem{helmstaedter2011high}
\bibinfo{author}{Helmstaedter, M.}, \bibinfo{author}{Briggman, K.~L.} \&
  \bibinfo{author}{Denk, W.}
\newblock \bibinfo{title}{High-accuracy neurite reconstruction for
  high-throughput neuroanatomy}.
\newblock \emph{\bibinfo{journal}{Nature {N}euroscience}}
  \textbf{\bibinfo{volume}{14}}, \bibinfo{pages}{1081--1088}
  (\bibinfo{year}{2011}).

\bibitem{roncal2015automated}
\bibinfo{author}{Gray~Roncal, W.} \emph{et~al.}
\newblock \bibinfo{title}{An automated images-to-graphs framework for high
  resolution connectomics}.
\newblock \emph{\bibinfo{journal}{Frontiers in {N}euroinformatics}}
  \textbf{\bibinfo{volume}{9}} (\bibinfo{year}{2015}).

\bibitem{Kasthuri2015}
\bibinfo{author}{Kasthuri, N.} \emph{et~al.}
\newblock \bibinfo{title}{Saturated reconstruction of a volume of neocortex}.
\newblock \emph{\bibinfo{journal}{Cell}} \textbf{\bibinfo{volume}{162}}
  (\bibinfo{year}{2015}).

\bibitem{mvrnch2009stripe}
\bibinfo{author}{M\"unch, B.}, \bibinfo{author}{Trtik, P.},
  \bibinfo{author}{Marone, F.} \& \bibinfo{author}{Stampanoni, M.}
\newblock \bibinfo{title}{Stripe and ring artifact removal with combined
  wavelet-fourier filtering}.
\newblock \emph{\bibinfo{journal}{Optics {E}xpress}}
  \textbf{\bibinfo{volume}{17}}, \bibinfo{pages}{8567--8591}
  (\bibinfo{year}{2009}).

\bibitem{donath2006automated}
\bibinfo{author}{Donath, T.}, \bibinfo{author}{Beckmann, F.} \&
  \bibinfo{author}{Schreyer, A.}
\newblock \bibinfo{title}{Automated determination of the center of rotation in
  tomography data}.
\newblock \emph{\bibinfo{journal}{JOSA A}} \textbf{\bibinfo{volume}{23}},
  \bibinfo{pages}{1048--1057} (\bibinfo{year}{2006}).

\bibitem{dowd1999developments}
\bibinfo{author}{Dowd, B.~A.} \emph{et~al.}
\newblock \bibinfo{title}{Developments in synchrotron {X}-ray computed
  microtomography at the national synchrotron light source}.
\newblock In \emph{\bibinfo{booktitle}{SPIE's International Symposium on
  Optical Science, Engineering, and Instrumentation}},
  \bibinfo{pages}{224--236} (\bibinfo{year}{1999}).

\bibitem{avinash1988principles}
\bibinfo{author}{Kak, A.~C.} \& \bibinfo{author}{Slaney, M.}
\newblock \emph{\bibinfo{title}{Principles of Computerized Tomographic
  Imaging}} (\bibinfo{publisher}{IEEE {P}ress}, \bibinfo{year}{1988}).

\bibitem{schindelin2012fiji}
\bibinfo{author}{Schindelin, J.} \emph{et~al.}
\newblock \bibinfo{title}{Fiji: an open-source platform for biological-image
  analysis}.
\newblock \emph{\bibinfo{journal}{Nature {M}ethods}}
  \textbf{\bibinfo{volume}{9}}, \bibinfo{pages}{676--682}
  (\bibinfo{year}{2012}).

\bibitem{burns2013open}
\bibinfo{author}{Burns, R.} \emph{et~al.}
\newblock \bibinfo{title}{The {O}pen {C}onnectome {P}roject data cluster:
  scalable analysis and vision for high-throughput neuroscience}.
\newblock In \emph{\bibinfo{booktitle}{Proc.\ of the 25th International
  Conference on Scientific and Statistical Database Management}},
  \bibinfo{pages}{27} (\bibinfo{organization}{ACM}, \bibinfo{year}{2013}).

\bibitem{davis1997adaptive}
\bibinfo{author}{Davis, G.}, \bibinfo{author}{Mallat, S.} \&
  \bibinfo{author}{Avellaneda, M.}
\newblock \bibinfo{title}{Adaptive greedy approximations}.
\newblock \emph{\bibinfo{journal}{Constructive Approximation}}
  \textbf{\bibinfo{volume}{13}}, \bibinfo{pages}{57--98}
  (\bibinfo{year}{1997}).

\end{thebibliography}
\bibliographystyle{nature}


\newpage
\section*{Supplemental materials}
\setcounter{figure}{0}  

\begin{figure}[h!]
\centering{
\includegraphics[width=0.85\textwidth]{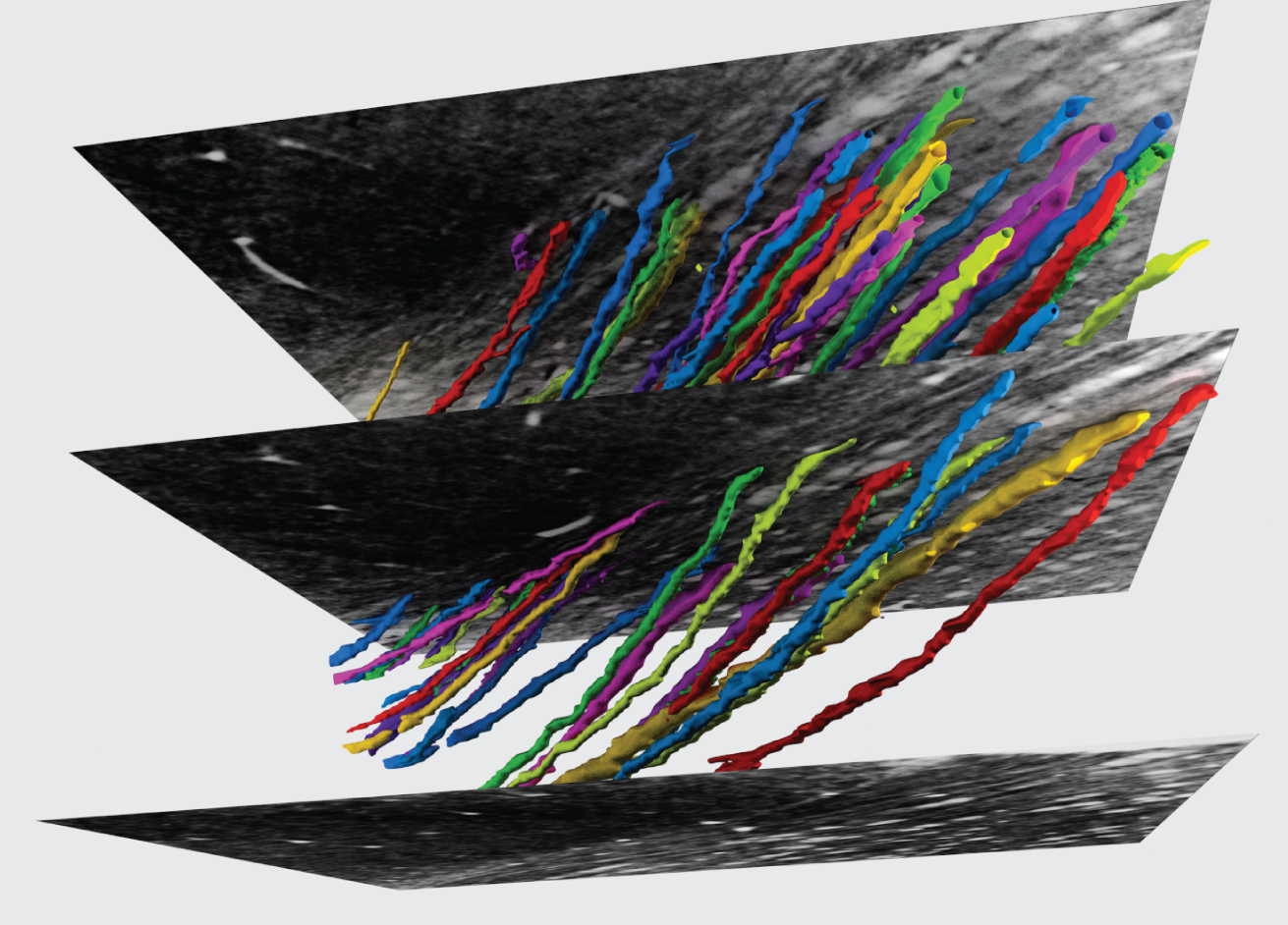}
\caption{\label{suppfig:axons} \small \linespread{0.1}{\em Manual reconstructions of myelinated axons.} A 3D visualization of myelinated axons, from mouse S1, after being manually traced in ITK-Snap. Many of these axons are only a few voxels in width, making them difficult to trace with automated algorithms.}}
\end{figure}

\begin{figure}
\centering{
\includegraphics[width=0.85\textwidth]{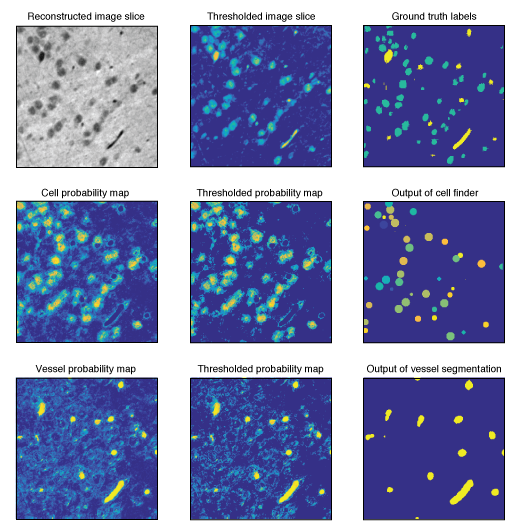}
\caption{\label{fig:algofig} \small \linespread{0.1}{\em Results of X-BRAIN pipeline for vessel segmentation and cell detection.}
In the top row, we display (left) a reconstructed image slice in false color, (middle) mean thresholded slice, and (right)  ground truth labels for both cells (green) and vessels (yellow). In the second row, we display: (left) the cell probability map we obtained after training a random forest classifier on the data with ilastik, (middle) the mean thresholded probability map, and (right) the output of our greedy sphere finder approach which operates on the cell probability map to obtain an estimate of the centroid and diameter of cells. In the third row along the bottom, we display: (left) the vessel probability map, (middle) the thresholded map, and (right) the output of our segmentation algorithm.
}}
\end{figure}

\begin{table*}[h!]
{
\label{tab:bigtable}
\begin{center}
\begin{tabular}{ccccc}
\toprule[1.5pt]
Annotation     	&  \# Cells 	&  Cell area & Volume ($\%$ of mm$^3$)	&  Cell density ($10^5/{\rm mm}^3$) 	 \\ \midrule

V0-A0	&	\quad  $97$	&	($2136,2060$)	&	$0.06$		&	$1.63$\\

V0-A1	&	  \quad  $96$	&	($1489,1499$)	&	$0.06$		&	$1.28$\\

V0-Xbrain	&	\quad $94$	&     ($1983,2123,51$)		& 	$0.06$	&	$1.57$\\
V1-A0	&	\quad $321$ 	&	($1997,2035$)	& 	$2.5$	& 	$1.28$ \\
V1-Xbrain	&	  \quad $302$	&  ($1983,1963,56$) &	$2.5$ 	&	$1.21$\\
%
%
V2-Xbrain	& \quad $112$ &   ($1918,1963,62$)	& 	$0.06$		&	$1.87$\\
V3-A03	&	  \quad  $281$	&      N/A	& 	$0.2$		&	$1.41$\\
V3-Xbrain	&	  \quad  $240$	&    ($1419,1385,42$)	& 	$0.2$		&	$1.20$\\
Vtot-Xbrain	&	$48,689$	&   ($1454, 1385,60$)	& $42$  &	$1.02$\\
\bottomrule[1.5pt]
\end{tabular}
\end{center}
\caption{{\em Statistics of manually labeled volumes, cell counts, and sizes for different volumes and annotators.} In the first column, we display the name of the volume (V0, V1, V2, and V3) and annotator to identify each manual (A0, A1, A2, A3) or automated (Xbrain) annotation. In the second column and third columns, we report the number of detected cells and the mean/median size of annotated cell bodies (number of labeled voxels). The training datasets include V0 (a subset of V1), V1, and V2. Volume V3 is held-out test set which whose location was unknown during training and tuning the parameters of the algorithm.}
}
\end{table*}

\begin{figure}
\centering{
\includegraphics[width=0.4\textwidth]{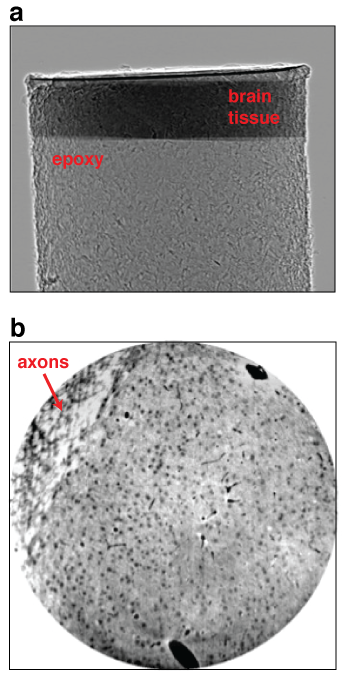}
\caption{\label{fig:suppfig} \small \linespread{0.1}{\em Data collected from beamline 32-ID.} (a) projection image with brain tissue embedded in epoxy and (b) slice reconstruction of the same sample acquired at 32-ID beamline using short-period undulator, with a single-line spectrum at 25 keV and a bandwidth of $10^{-2}$. This type of undulator enables radiograph acquisition without any optics between the source and the sample, leading to relatively artifact-free flat-field corrected images (a). Synchrotron sources that use undulators produce a smaller source with a larger coherence length and thus improve sensitivity when coupled with a propagation-based phase contrast approach. In the image (b), we can clearly resolve a bundle of myelinated axons in the sample.
}}
\end{figure}

\begin{figure}
\centering{
\includegraphics[width=0.9\textwidth]{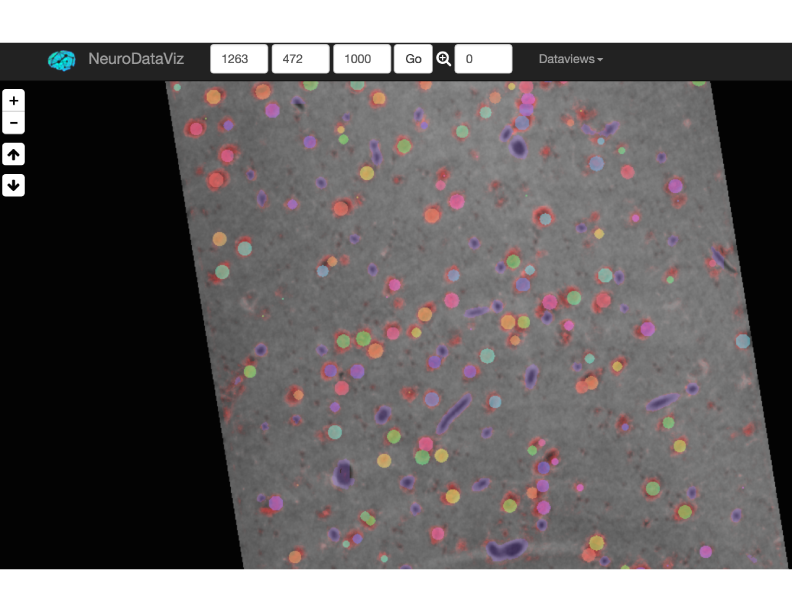}
\caption{\label{fig:suppfig2} \small \linespread{0.1}{\em Visualization of cell and vessel segmentation performance.} The results of X-BRAIN processing on our data sample as visualized through NeuroData's visualization service (ndviz). This view can be found here (\protect\url{viz.neurodata.io/project/xbrain/0/1263/472/1000}) and users can easily traverse through the volume using NeuroData's web-based GUI. The cell probabilities (translucent red) and final cell detections (opaque multi-color, where each color represents a unique ID for a cell), and the vessel segmentation (translucent purple), are all overlaid on the corresponding X-ray image.}}
\end{figure}

\end{document}